\documentstyle[12pt]{article}
\input psfig

\def\reference{\parskip 0pt\par\noindent\hangindent 0.5 truecm}

\def\sun{\odot}

\begin{document}

\title{On the Gas Surrounding High Redshift Galaxy 
Clusters$^1$}
\footnotetext[1]{Based on observations carried out at the Anglo Australian
Telescope, Cerro Tololo Interamerican Observatory and at Siding Spring
Observatory.}

\author{Paul J Francis $^{2,3}$ \and
 Greg M. Wilson $^{2}$ \and
Bruce E. Woodgate $^{4}$
}

\date{}
\maketitle

{\center
$^2$ Research School of Astronomy and Astrophysics, Australian National
University, Canberra ACT 0200: pfrancis, gmw@mso.anu.edu.au\\[3mm]
$^3$ Joint appointment with the Department of Physics, Faculty of Science\\[3mm]
$^4$ NASA Goddard Space-Flight Center, Code 681, Greenbelt, MD 20771, USA: 
woodgate@s2.gsfc.nasa.gov \\
[3mm]
}

%
\begin{abstract}

Francis \& Hewett (1993) identified two 10-Mpc scale regions of the
high redshift universe that were seemingly very overdense in neutral
hydrogen. Subsequent observations showed that at least
one of these gas-rich regions enveloped a cluster of galaxies at
redshift 2.38. We present improved observations of the three background QSOs 
with sightlines passing within a few Mpc of this cluster of galaxies.
All three QSOs show strong neutral hydrogen absorption at the cluster redshift,
suggesting that this cluster (and perhaps all high redshift clusters) may
be surrounded by a $\sim 5$ Mpc scale region containing 
$\sim 10^{12} M_{\sun}$ of neutral gas.

If most high redshift clusters are surrounded by such regions,
we show that the gas must be in the form of many small ($< 1$ kpc),
dense ($> 0.03 {\rm \ cm}^{-3}$) clouds, each of mass $< 10^6 M_{\sun}$.
These clouds are themselves probably gathered into $> 20$ kpc sized clumps,
which may be galaxy halos or protogalaxies.

If this gas exists, it will be partially photoionised by the UV background.
We predict the diffuse Ly$\alpha$ flux from this photoionisation, and place
observational limits on its intensity.

\end{abstract}

\bigskip

\section{Introduction}

What did galaxy clusters look like ten billion years ago? 
Simulations (eg. Brainerd \& Villumsen 1994, Jenkins et al. 1998, 
Cen 1998) suggest that the ancestors of 
present day rich galaxy clusters contained very little mass at redshifts
above two: there simply hadn't been time enough to assemble much dark or 
baryonic
matter. Even the richest proto-clusters would only have had about double
the average density of the universe. These proto-clusters would typically
have been filamentary in shape, and due to their small masses would not
have been in virial equilibrium.

Despite their modest overdensities, these protoclusters could have
been the sites of the first galaxy formation. If galaxies had formed
in these proto-clusters but not in the field, the overdensity of
galaxies within the proto-cluster would be much greater than the
overdensity of mass (biasing, eg. Fry 1996, Bagla 1998, Baugh et al. 1999,
Tegmark \& Peebles 1998).

The observational situation is less clear. Galaxy clusters are now routinely
being studied out to redshifts $z \sim 1$. These clusters are remarkably
similar to low redshift clusters: they are massive virially bound objects
with strong X-ray emission (eg. Rosati et al. 1998, Deltorn et al. 1997). 
Thus galaxy clusters 
appear to be well established by redshift one (see also Renzini 1997, Donahue 
\& Voit 1999).

Observations at higher redshifts still are sparse. It is now clear that
galaxies at $z>2$ are at least as strongly clustered as galaxies
today (eg. Heisler, Hogan \& White 1989, Quashnock, Vanden Berk \& York 
1996, Malkan, Teplitz \& McLean 1996, Steidel et al. 1998, 
Giavalisco et al. 1998, Campos et al. 1999, Pascarelle, Windhorst 
\& Keel 1998, Adelberger et al. 1998, Djorgovski et al. 1999), as 
predicted by the biasing model. The structure, mass and galaxy populations 
of these galaxy concentrations are, however, obscure.

A growing body of evidence suggests that high redshift clusters contain
substantial quantities of gas. The strongest evidence comes from high
redshift radio galaxies, which are believed to sit in dense cluster
environments  (eg. Pentericci et al. 1997, Ivison et al. 2000). Many show 
high rotation measures (Carilli et al. 1997, Pentericci et al. 2000), 
possibly extended X-ray emission (Carilli et al.
1998), extensive emission-line halos (Bicknell et al. 2000) and 
associated absorption lines (Binette et al. 2000), suggesting the presence of 
dense inhomogeneous gas around the galaxies.

Are the clusters around radio galaxies typical? The evidence for gas in
other high redshift clusters is sparse. Francis et al. (1996) and Steidel 
et al. (1998) both found QSO absorption lines seemingly coincident with 
high-z clusters. Tentative Sunyaev-Zel'dovich effect decrement measurements 
have been claimed around other possible high-z clusters (eg. Campos et al. 
1999). 100 Kpc scale diffuse
`blobs' of Ly$\alpha$ emission have been found in the centres of two high-z
clusters (Francis et al. 1996, Francis, Woodgate \& Danks 1997, 
Steidel et al. 2000).

Most of these observations suggest the presence of hot ionised gas in the 
central few hundred kpc of the clusters. Francis \& Hewett (1993), however,
presented tentative evidence for the existence of large concentrations of 
neutral gas extending over Mpc scales in the high redshift universe. One of 
their concentrations was associated with a group of galaxies (Francis,
Woodgate \& Danks 1997).
Their evidence was not conclusive, and there is currently no such data 
available for other clusters, but it is at least possible that similar
extended halos of neutral gas surround all high redshift clusters.

In this paper, we concentrate on one of Francis \& Hewett's clusters:
the 2142$-$4420 galaxy concentration at redshift 2.38. In Section~\ref{cluster}
we present new data on this cluster, strengthening the case for a large 
halo of neutral gas surrounding it.
In Section~\ref{dist} we investigate ways of reconciling high neutral
gas masses with cosmological models. In Section~\ref{diffuse} we describe
our attempt to observe diffuse Ly$\alpha$ emission from the hypothesised gas
halo, and place upper limits on its intensity.
Except where stated, we
assume $H_0 = 70 {\rm km\ s}^{-1}{\rm Mpc}^{-1}$, and an open universe with 
$\Omega_0 = 0.2$ and $\Lambda =$ either $ 0$ or $0.8$ (both values of
$\Lambda$ give the same angular size distance and luminosity distance at
redshift 2.38).

\section{Observations of the 2142$-$4420 Cluster\label{cluster}}

\subsection{Review of the 2142$-$4420 Cluster Properties\label{review}}

The 2142$-$4420 cluster lies at redshift 2.38, at coordinates 
21:42:30$-$44:20:30 (J2000). As will be shown, it is a region of the 
early universe that is highly overdense in Ly$\alpha$ emitting galaxies.
It is unlikely to be gravitationally bound, and so would not meet most low
redshift definitions of a galaxy cluster. 

Francis \& Hewett originally identified the cluster as a pair of strong Lyman
limit systems at matching redshifts (2.38) in two 19th magnitude QSOs:
2138$-$4427 and 2139$-$4434. The QSO sight-lines are 500$^{\prime \prime}$
($\sim 4$ proper Mpc) apart at $z=2.38$ (both QSOs lie at $z \sim 3.2$). Such 
a pair of Lyman-limit 
systems at matching wavelengths is unlikely to occur by chance in Francis 
\& Hewett's sample. As we report in Section~\ref{3spec}, a third QSO has
been found behind the cluster, and it too shows strong Ly$\alpha$
absorption at the cluster redshift.

Are there any galaxies associated with this concentration of QSO absorption
lines? Many high redshift galaxies show {\em weak} Ly$\alpha$ emission 
(Steidel et al. 1996, Hu, Cowie \& McMahon 1998). The field containing
the three QSOs, however, contains three {\em strong} Ly$\alpha$ emitting 
galaxies at $z=2.38$, with Ly$\alpha$ fluxes $F_{\rm Ly \alpha} > 
10^{-16}{\rm erg\ cm}^{-2}{\rm s}^{-1}$ (Francis et al. 1996, 
Francis, Woodgate \& Danks 1997); an order of
magnitude greater than that of normal galaxies at these redshifts. All three 
have reliable spectroscopic redshifts.

Do three such sources constitute a cluster? Strong Ly$\alpha$ 
emitters are rare at these redshifts: Francis et al. surveyed a total volume 
of 460 co-moving cubic Mpc,  
but all three sources were found within a 5 cubic co-moving Mpc volume: 
ie. $\sim$ 1\% of the surveyed co-moving volume. Mart\'{\i}nez-Gonz\'alez 
et al. (1995) surveyed a co-moving volume of 1400 cubic
Mpc at $z=3.4$ for Ly$\alpha$ emitting sources to a comparable flux limit 
but detected nothing. The odds of the cluster being an artifact of the
coincidental proximity of three such sources is thus $< (5/1860)^2$
(the probability of finding two more such objects within one proper Mpc of the 
first), ie. $< 10^{-5}$. Clearly the space density of Ly$\alpha$ emitting 
galaxies in this region is higher than average.

Could the presence of absorption in all three QSOs be coincidental, or 
is the cluster really surrounded by Lyman-limit absorption-line systems?
The transverse separation of the QSO sight-lines is $\sim 5$ Mpc, which
corresponds to a redshift difference of $\sim 20$\AA\ along the
line of sight. The absorption-line systems all have equivalent widths of
$> 20$\AA . The probability of seeing an absorption line with an equivalent
width this strong within $\pm 20$\AA\ of any given wavelength is $1.3$\%
(Francis \& Hewett 1993). Thus the probability of finding three such absorption-line
systems within the cluster by chance is $2 \times 10^{-6}$.

This calculation should be regarded with caution: this region was first
identified as interesting because of the absorption in the two original
QSOs (Francis \& Hewett 1993), so the statistics are a posteriori. The 
third QSO,
however, was not involved in selecting this region for study, and its
coincident absorption alone makes this region overdense in Lyman-limit
systems with 98\% confidence.

\begin{figure}
\psfig{file=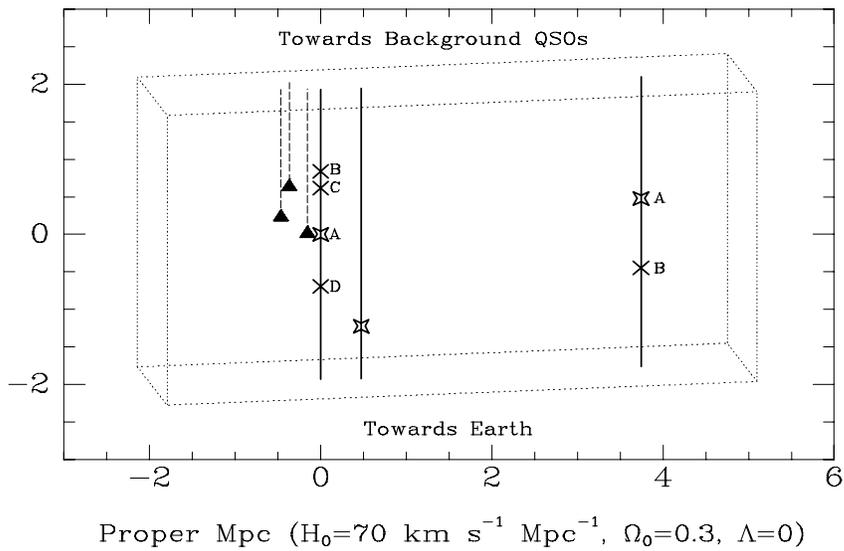,height=100mm}
\caption{3D view of the 2142$-$4420 cluster. All objects have been 
assigned 3-dimensional
positions, based on their location on the sky and their redshift, assuming
that all redshifts trace the Hubble Flow. Solid lines are the sight-lines to 
the three background QSOs. Stars are Ly-limit absorption systems, crosses are
the lower column density absorption systems, and solid triangles are the
Ly$\alpha$ emitting galaxies. Redshifts increase upwards: galaxy redshifts
have been derived, where possible, from the metal emission 
lines. From left to right, the QSO sight lines are 2139$-$4434,
2139$-$4433 and 2138$-$4427.\label{threed}}

\end{figure}

Fig.~\ref{threed} should make the geometry of the cluster clearer. Due to
the low predicted overdensities of clusters at this redshift 
(Section~\ref{mass}), peculiar motions should be very small, so three 
dimensional 
positions are plotted assuming that all redshift differences are due to 
distance. The three Ly$\alpha$ galaxies lie within one Mpc of each other. 
The absorption-line systems are far more dispersed, extending both 
to lower redshifts and transversely by $\sim 5$ Mpc.

The spatial extent and overdensity of this cluster are comparable
to those of the clusters of Lyman-break galaxies being found by 
Steidel et al (1998) at $3 < z < 3.5$: this cluster may be a 
representative of the same class of object.

\subsection{Observations\label{3spec}}

\begin{figure}
\psfig{file=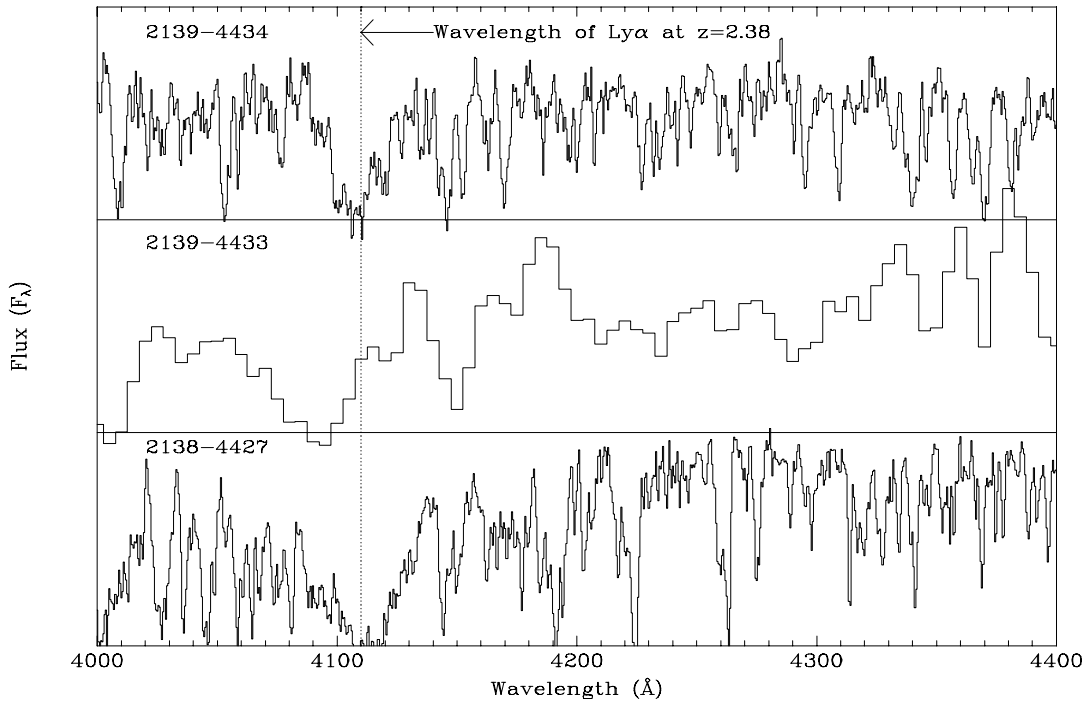,height=100mm}
\caption{Spectra of the three background QSOs, showing 
the Ly$\alpha$ 
absorption at the cluster redshift (4108 \AA ). The top and bottom
spectra are the original AAT spectra: the middle panel is the new LDSS
spectrum.\label{threeabs}}
\end{figure}

V\'eron \& Hawkins (1995) searched an area including
this cluster for variable sources. In addition to both previously 
identified QSOs, they discovered a third QSO lying between the two:
QSO 2139$-$4433 at z=3.22 (ie. at the same redshift as the other two
background QSOs). We measured a position for QSO 2139$-$4433 
(21:42:22.16$-$44:19:28.7, J2000) using our R-band image with an astrometric 
solution bootstrapped from on-line scans of UK Schmidt plates 
(Drinkwater, Barnes \& Ellison 1995). A spectrum was obtained with the 
Low Dispersion Survey 
Spectrograph (LDSS, Colless et al. 1990) on the Anglo-Australian 
Telescope on the 
nights of 1996 August 13 and 14. The total exposure time was 47,700 sec, and 
the spectral resolution 700${\rm km\ s}^{-1}$. Part of the spectrum is shown 
in Fig~\ref{threeabs}.

As Fig~\ref{threeabs} shows, the new QSO 2139$-$4433 has a strong
absorption-line system close in wavelength to the absorption in the
two previously known QSOs at z=2.38. This further confirms the remarkable
gas properties of this cluster.

Our original spectra of QSOs LBQS 2138$-$4427 and 2139$-$4434 are described by
Francis \& Hewett (1993). Their resolution was excellent
(full width at half maximum height $100 {\rm km\ s}^{-1}$) but the wavelength 
coverage (4000 -- 4600 \AA ) was
small. An additional spectrum of QSO 2139$-$4434 was obtained with the
KPGL1 grating in the Blue Air camera of the RC spectrograph on the 
CTIO 4-m telescope on 1995 August 20. Total
exposure time was 12,000 sec, with a spectral resolution of 
200${\rm km\ s}^{-1}$. This spectrum, while inferior in resolution to the
spectrum of Francis \& Hewett, covers 3200 -- 6200 \AA : this greater 
wavelength
range allows us to study CIV and Ly$\beta$ absorption from the cluster.

\begin{figure}
\psfig{file=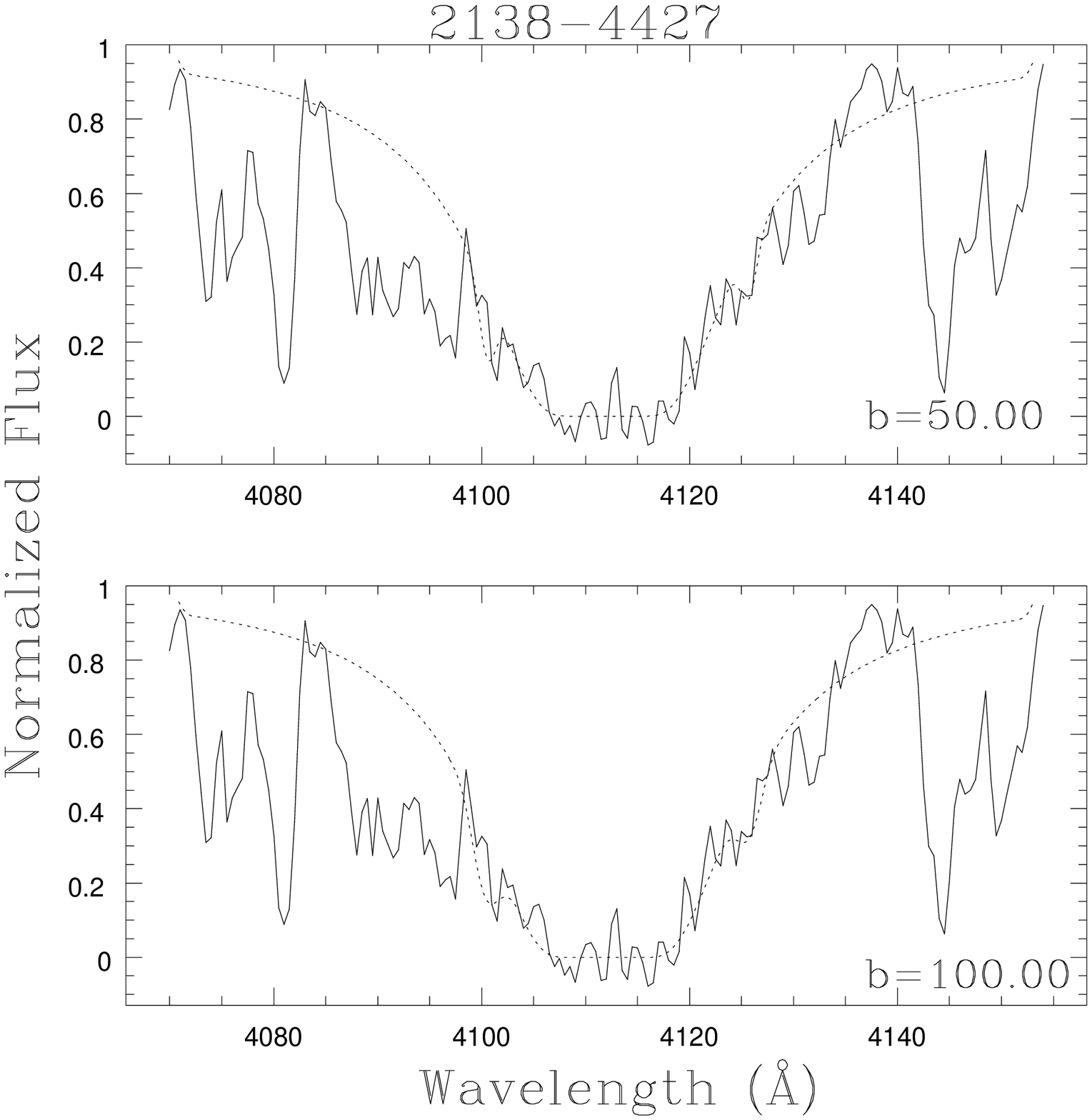,height=100mm}
\psfig{file=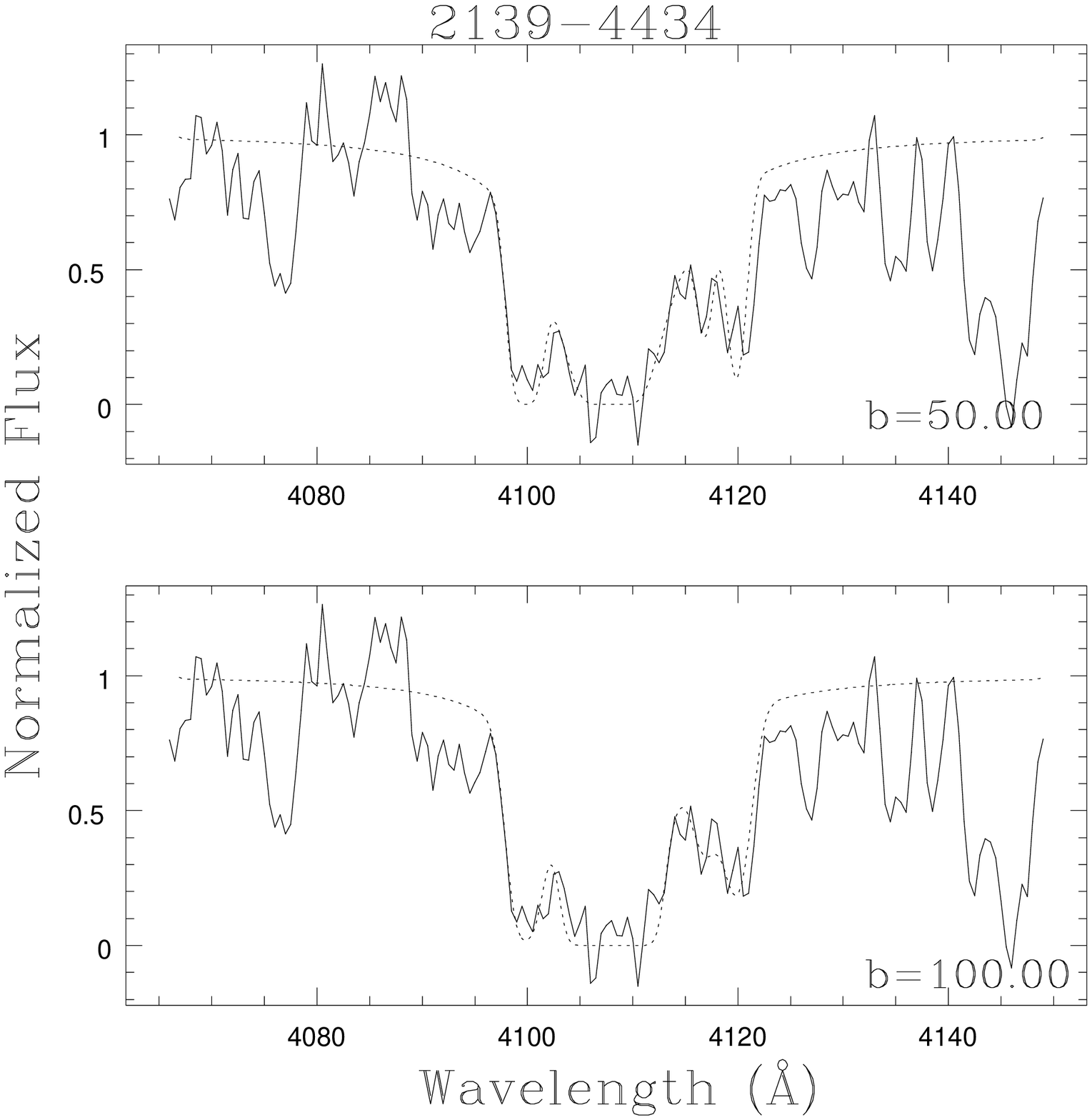,height=100mm}
\caption{Voigt profile fits to the Ly$\alpha$ 
absorption at the cluster
redshift in QSOs 2138$-$4427 and 2139$-$4434. Fits are shown for the
two different velocity dispersions $b$ assumed. \label{lyfits}}
\end{figure}

\subsection{Absorption Line Measurements}

Combining the old and new data on the two brighter QSOs, we fit Voigt 
profiles interactively to the absorption at the cluster redshift, 
using the Xvoigt program (Mar \& Bailey 1995). The low spectral resolution, 
restricted
wavelength coverage and blending in our spectra make this process a
difficult and ambiguous one. Nonetheless, certain definite conclusions
can be reached. Multiple components are required to obtain adequate fits 
to the Ly$\alpha$ absorption (Fig~\ref{lyfits}). A minimum of 2--4 components 
are required (Tables~\ref{abs38}, \ref{abs39}): many more, each with smaller
column densities, give equally good fits. The column densities of the
subsidiary systems are not well constrained. We could not determine the 
velocity dispersion $b$ of the metal lines: an upper limit of $\sim 100 
{\rm km\ s}^{-1}$ can be placed. The central component of the 
Ly$\alpha$ absorption in all three QSOs was broader: the flux touches zero 
over $\sim 100 {\rm km\ s}^{-1}$ or more.

\begin{table}
\begin{center}
\caption{Possible Absorption Systems in QSO 2138$-$4427 \label{abs38}}
\begin{tabular}{llccc}
\hline
\multicolumn{3}{c}{ ~ } &
\multicolumn{2}{c}{Column density $\log(N[cm^{-2}])$} \\
System & 
Ion & 
Redshift & 
$b=50 {\rm km\ s}^{-1}$ &
$b=100 {\rm km\ s}^{-1}$ \\
\hline \\
A & H I  121.6 & 2.3825 & 20.47 & 20.47 \\
  & Si II 130.4 & 2.3823 & 14.63 & 14.72 \\
  & Si II 126.0 & 2.3825 & 14.63 & 14.22 \\
  & Si III 120.6 & 2.3824 & 14.30 & 13.98 \\
  & O I  130.2 & 2.3820 & 16.22 & 15.47 \\
  & C I  127.7 & 2.3821 & 14.62 & 14.70 \\
  & C II  133.4 & 2.3822 & 16.00 & 15.13 \\
B & H I  121.6 & 2.3731 & 13.97 & 14.10 \\
\hline
\end{tabular}
\end{center}
\end{table}

\begin{table}
\begin{center}
\caption{Possible Absorption Systems in QSO 2139$-$4434 \label{abs39}}
\begin{tabular}{llccc}
\hline
\multicolumn{3}{c}{ ~ } &
\multicolumn{2}{c}{Column density $\log(N[cm^{-2}])$} \\
System & 
Ion & 
Redshift & 
$b=50 {\rm km\ s}^{-1}$ &
$b=100 {\rm km\ s}^{-1}$ \\
\hline \\
A & H I  121.6 & 2.3792 & 19.80 & 19.67 \\
  & Si II 130.4 & 2.3794 & 13.63 & 13.80 \\
  & Si II 126.0 & 2.3792 & 13.53 & 13.67 \\
  & Si III 120.6 & 2.3789 & 13.45 & 13.55 \\
  & O I  130.2 & 2.3804 & 16.00 & 15.47 \\
  & C I  127.7 & 2.3790 & 14.00 & 14.17 \\
  & C II  133.4 & 2.3783 & 16.03 & 15.13 \\
  & C IV  154.8 & 2.3787 & 14.30 & 14.23 \\
  & C IV  155.1 & 2.3787 & 14.17 & 14.43 \\
B & H I  121.6 & 2.3890 & 14.56 & 14.37 \\
C & H I  121.6 & 2.3865 & 14.07 & 14.07 \\
D & H I  121.6 & 2.3724 & 16.93 & 14.80 \\
\hline
\end{tabular}
\end{center}
\end{table}

We searched for metal-line absorption at the redshift of the dominant 
Ly$\alpha$ absorption components. With the exception of C IV, these
all lie within the Ly$\alpha$ forest, and hence may be chance coincidences
with the forest lines. The strongest line near the expected wavelength
was fit, assuming velocity widths of $50$ and $100 {\rm km\ s}^{-1}$, and
the results are shown in Tables~\ref{abs38} and \ref{abs39}. Due to the risk
of blending or confusion with Ly$\alpha$ forest lines, the metal line
column densities should be taken as upper limits. Plots of the metal lines
within the forest can be found in Francis \& Hewett. Note that
in QSO 2138$-$4427, strong absorption lines were invariably detected at
the expected wavelengths, while in QSO 2139$-$4434 the lines were weaker
and at slightly shifted wavelengths. We conclude that the central 
absorption component in QSO 2138$-$4427 does contain metals, roughly
as measured, while for QSO 2139$-$4434 some or all of the putative
lines (except C IV) may be misidentified Ly$\alpha$ forest lines.
Our spectrum of QSO 2139$-$4433 had too low a resolution
to determine anything other than the Ly$\alpha$ redshift (2.366) and
column density ($\log(N_H) \sim 20.7$).

Are the absorption-line systems really Lyman-limit systems, or could they
just be clusters of lower column density Ly$\alpha$ forest lines? 
In QSO 2138$-$4427 the Ly$\alpha$ line shows broad wings, and strong 
absorption is seen at the expected wavelength of most common metal
absorption lines: it therefore seems probable that this is, as modelled, 
a high column density absorption system, probably lying on the column density
borderline between Lyman-limit and damped Ly$\alpha$ systems. The spectrum 
of QSO 2139$-$4433 is of too low resolution to say much, but the great width 
and equivalent width of the Ly$\alpha$ absorption also suggest that its
absorption column is large.

In QSO 2139$-$4434, however, the situation is more ambiguous. It is 
possible to fit the Ly$\alpha$ absorption either with a single absorption-line
system with column density $N_H \sim 10^{19.7}{\rm cm}^{-2}$ (plus three 
much weaker components in the wings), or 
with a blend of weaker Ly$\alpha$ lines, spread over $\sim 200 {\rm km\ 
s}^{-1}$ and with a combined neutral hydrogen column density that can be
as low as $N_H \sim 10^{16.5}{\rm cm}^{-2}$. Two pieces of evidence support 
this latter fit. Firstly, the redshifts of the
supposed metal-lines vary by $\pm 80 {\rm km \ s}^{-1}$ (though some or all may
be chance coincidences with Ly$\alpha$ forest lines). This can be explained
if they are coming from different subcomponents of the absorption system.
Secondly, there is
tentative evidence that Ly$\beta$ absorption is weak: the spectrum is
poor at this wavelength, and the continuum hard to define, but the Ly$\beta$
absorption can be well fit with column densities as low as $N_H 
\sim 10^{16}$ (though much greater columns also give acceptable fits). 
On the other hand, the strength of the metal lines, especially low
ionisation lines such as C II, imply that the neutral hydrogen 
column density is $N_H > 10^{18}{\rm cm}^{-2}$. Note, however, that with 
the exception of C IV, these lines could be contaminated by Ly$\alpha$
forest absorption.

Note that the total gas column density
in the form of the absorbing clouds is almost independent of the
interpretation of the data. Gas with a {\em neutral} column density of $\sim
10^{19}{\rm cm}^{-2}$ is predicted to be mostly neutral and hence to have a 
{\em total} gas column density $\sim 10^{19}{\rm cm}^{-2}$. Gas with a 
{\em neutral} column
density of $\sim 10^{16}{\rm cm}^{-2}$, on the other hand, is predicted to
be strongly ionised by the UV background, and hence its {\em total} gaseous
column density will be $\sim \times 10^3$ greater than the neutral column
density. Thus the total hydrogen column would be roughly the same as for 
the damped Ly$\alpha$ interpretation. 

\section{The Nature of the Gas\label{dist}}

The three QSO sight-lines may be unrepresentative of the space
around this cluster. If they are representative, however, this implies
that a fraction of order unity of all sight-lines passing within several Mpc
of this cluster would intersect a cloud of hydrogen with a neutral column 
density of $N_H \sim 10^{19} {\rm cm}^{-2}$ or greater.

In this section we hypothesise that the region surrounding the 
2142$-$4420 cluster is optically
thick in neutral hydrogen clouds. We further hypothesise that this cluster
is typical of clusters at this redshift. What would be the  physical
consequences if our hypotheses were correct?

\subsection{Gas Geometry}

The absorption-line gas in cluster 2142$ - $4420 seems to be substantially 
more extended spatially
than the Ly $\alpha$ emitting galaxies, as can be seen in Fig~\ref{threed}.
Strong absorption is seen in QSO 2138$-$4427, whose sight-line passes
$\sim 4$ Mpc from the concentration of Ly$\alpha$ emitting galaxies. It is
also seen in QSO 2139$-$4433, at a redshift that places the absorption
about 2 Mpc in front of the cluster. We therefore hypothesise that the
cluster of Ly$\alpha$ sources is embedded within a much larger
structure of absorbing gas.
Indeed, it is suggestive that the QSO sight-line passing closest to the
three Ly$\alpha$ emitting galaxies shows the lowest neutral hydrogen column 
density. This might indicate that the central region of the neutral gas
structure is hotter and more ionised than the outer regions.

If this neutral gas structure does exist, what is its geometry? With only 
the QSO absorption to guide us, all we can do is bracket the possibilities 
with three straw models:

\begin{enumerate}

\item Sheet Model: the gas lies in a sheet, of thickness $\sim 1$ Mpc,
width $> 4$ Mpc and depth (along the line of sight) $\sim 4$ Mpc. This
sheet would be edge-on to our sight-line. 

\item Spherical model: The gas lies in a spherical halo, centred on the
 Ly$\alpha$ emitting galaxies. The halo radius must be at least 4 Mpc.

\item  Filamentary Model: the cluster of galaxies lie at the intersection
of a number of gas-filled filaments. Each filament is at least 4 Mpc long,
and perhaps around 1 Mpc thick. Filamentary distributions are
predicted by many cosmological simulations (eg. Rauch, Haehnelt \& 
Steinmetz 1997).

\end{enumerate}

Is the region around the galaxy cluster completely full of absorbing gas? Let 
us assume that a fraction $f$ of all randomly chosen site-lines passing 
within $\sim 5$ proper Mpc of the galaxy cluster  
would intercept at least one gas cloud with a total hydrogen cross
section $>10^{19}{\rm cm}^{-2}$. The probability of all three of our
QSO sight-lines showing such absorption is thus $f^3$. If we require
this probability to be greater than 1\%, that implies $f> 22$\%.

\subsection{The Mass and Density of the Neutral Gas Structure\label{mass}}

Let us assume that the cluster really is surrounded by a $\sim 5$  proper Mpc
scale structure of gas with $f \sim 1$ and a typical neutral absorption-line
column density of $N_H \sim 10^{19.5}{\rm cm}^{-2}$. What would be the
consequences? The average density 
of absorption-line gas within the gas structure will then be  $\sim 8 \times
10^{10} M_{\sun}{\rm Mpc}^{-3}$ in proper coordinates. This is $\sim 100$
times greater than the typical density of neutral hydrogen at this
redshift (eg. Steidel 1990). 

Given this density, the combined gas mass of all the absorption-line 
systems within the neutral gas structure would be 
$> 6 \times 10^{11} M_{\sun}$ for the sheet or filamentary models, and
 $> 3 \times 10^{12} M_{\sun}$ for the spherical model.

This overdensity of baryonic matter in the form of hydrogen 
clouds could be explained in two ways:

\begin{enumerate}

\item The volume surrounding the galaxy cluster is overdense in all forms of 
matter: the 
efficiency with which baryons form absorbing gas clouds and red galaxies
is the same as elsewhere in the universe. 

\item The volume is not greatly overdense, but the efficiency with
which baryons formed neutral gas clouds and red galaxies is enhanced in
this region.

\end{enumerate}

Cosmological simulations favour the second explanation.
At redshift 2.38, there has not been sufficient time to
assemble large matter concentrations. Many authors have used analytic
methods or n-body simulations to estimate the mass of matter concentrations
in the high redshift universe. In Table~\ref{xitable}, we have collected
a sample of these results, parameterised as the two-point correlation 
coefficients $\xi$ for mass on the approximate scale of the gas structure, at 
redshifts $z \sim 2.38$.

\begin{table}
\begin{center}
\caption{Predicted mass two-point correlation coefficients 
$\xi$. \label{xitable}}
\begin{tabular}{clccc}
\hline
Ref. & Model Details & Co-moving Radius & z & $\xi(r_0, z)$ \\
\hline \\
1 & SCDM: $\Omega_0=1.0$, $\Lambda=0$, $h=0.5$ (small $\sigma_8$) & 
$8.81h^{-1}$ Mpc & 2.3  & 0.06 \\ 
1 & SCDM: $\Omega_0=1.0$, $\Lambda=0$, $h=0.5$ (large $\sigma_8$) & 
$8.81h^{-1}$ Mpc & 2.3  & 0.17 \\ 
2 & $\Omega_0=1.0$, $\Lambda=0.0$ & 11.4 Mpc & 2.8 & 0.008 \\
2 & $\Omega_0=0.2$, $\Lambda=0.0$ & 11.4 Mpc & 2.8 & 0.095 \\ 
2 & $\Omega_0=0.2$, $\Lambda=0.8$ & 11.4 Mpc & 2.8 & 0.07  \\
3 & $\Omega_0=1.0$, $\Lambda=0.0$ & $10h^{-1}$ Mpc & 2.0 & 0.020  \\ 
3 & $\Omega_0=1.0$, $\Lambda=0.0$ & $10h^{-1}$ Mpc & 3.0 & 0.016  \\ 
4 & SCDM (with Zel'dovich Approximation) & $10h^{-1}$ Mpc & 2.0 & 0.05 \\
5 & SCDM: $\Omega_0=1$, $\Lambda=0$, $h=0.5$ (small $\sigma_10$) & $10h^{-1}$
Mpc & 2.4 & 0.0036 \\
5 & SCDM: $\Omega_0=1$, $\Lambda=0$, $h=0.5$ (large $\sigma_10$) & $10h^{-1}$
Mpc & 2.4 & 0.03 \\
5 & TCDM: $\Omega_0=1$, $\Lambda=0$, $h=0.5$ (small $\sigma_10$) & $10h^{-1}$
Mpc & 2.4 & 0.0014 \\
5 & TCDM: $\Omega_0=1$, $\Lambda=0$, $h=0.5$ (small $\sigma_10$) & $10h^{-1}$
Mpc & 2.4 & 0.0069 \\
5 & OCDM: $\Omega_0=0.4$, $\Lambda=0$, $h=0.65$ & 
$10h^{-1}$Mpc & 2.4 & 0.016 \\
5 & $\Lambda$CDM: $\Omega_0=0.4$, $\Lambda=0.6$, $h=0.65$  & $10h^{-1}$
Mpc & 2 & 0.0036 \\
6 & SCDM: $h=0.5$  & $10h^{-1}$ Mpc & 2.4 & 0.02 \\
7 & LCDM:  & $10h^{-1}$ Mpc & 3 & $<0.005$ \\
7 & C+HDM:  & $10h^{-1}$ Mpc & 3 & $<0.005$ \\
\hline 
\end{tabular}
\end{center}

\noindent
$[1]$ Brainerd \& Villumsen (1994). 

\noindent
$[2]$ C\'olin, Carlberg \& Couchman (1997).

\noindent
$[3]$ Matarrese et al. (1997).

\noindent
$[4]$ Porciani (1997).

\noindent
$[5]$ Moscardini et al. (1998).

\noindent
$[6]$ Bagla (1998).

\noindent
$[7]$ Ma (1999).

\end{table}

Given these values of $\xi$, we can estimate the root-mean squared (rms) 
fluctuations $\sigma_R$ between the average densities of various spherical 
regions of this size in the early universe. We use the approximate 
relationship 
\begin{equation}
\sigma_R^2\approx 2.5 \xi(R) \label{sigma},
\end{equation}
which is obtained from the relation $\sigma^2_R=3J_3(R)/R^3$ where a top hat
window function has been assumed and $J_3$ represents the integrated two 
point correlation function (Kolb \& Turner 1990). Thus 
\begin{equation}
J_3=\frac{1}{4\pi}\int_0^R\xi(r)d^3r,
\end{equation}
where the two point correlation function is assumed to be of the form
$\xi(r)=(r/r_0)^{-\gamma}$, with $\gamma\approx 1.8$ (Groth \& Peebles 1977, 
Davis \& Peebles 1983). 

Thus typical predicted rms mass fluctuations on the scale of this 
absorption-line structure ($\sim 10$ co-moving Mpc) are only $\sim 25\%$. Even 
for models with the
most extreme fluctuations (low density models: Cen 1998), and 
assuming that the cluster sits in a $5 \sigma$
mass fluctuation, the average density on this scale cannot be more than
twice the mean density of the universe. Note that this applies to
a roughly spherical volume: if the structure really
is sheet-like or filamentary, the overdensity {\em within this structure}
can be significantly greater.
  
So, the average mass density of the $\sim 10$ co-moving Mpc scale 
volume including the cluster must be of the same order as 
that of the universe as a whole at this redshift. If, as seems
likely, baryonic and non-baryonic matter trace each other on these large
scales, primordial nucleosynthesis thus gives us an approximate upper
limit on the average baryon density of the cluster halo. Assuming
$\Omega_{\rm baryon} = 0.016 h_{100}^{-2}$ (eg. Walker et al. 1991), and 
choosing $H_0 = 70 {\rm km\ s}^{-1}{\rm Mpc}^{-1}$,
this density is $1.2 \times 10^{-26} {\rm kg\ m}^{-3}$
($\sim 2 \times 10^{11} M_{\sun}{\rm Mpc}^{-3}$). 

Thus even if the QSO absorption-line column densities are representative
of the whole cluster halo, the baryonic mass of the cluster inferred
(0.6---3$ \times 10^{12}  M_{\sun}$ in the form of absorbing clouds) is
physically possible: it does not exceed the predicted baryonic mass within
the cluster volume ($\sim 7 \times 10^{12} M_{\sun}$). The puzzle would be the
high fraction of these baryons that are incorporated into absorption-line 
systems: the efficiency of formation of
these objects is at least 10\% and may well be much higher.
This efficiency is far higher than is typical at this or
any other redshift (eg. Cen \& Ostriker 1999)

\subsection{The Physical State of the Gas\label{gasphys}}

If the QSO sight-lines are representative of all sight-lines through the
region around the cluster, it seems to be embedded in a structure of size 
$\sim 5$ Mpc,
and most sight-lines through this structure intersect a gas cloud with a
hydrogen column density $N_H > 10^{19}{\rm cm}^{-2}$. Is there any
physically plausible structure with these properties?

\subsubsection{Argument from the Total Baryon Density\label{baryon}}

In this section, it is shown that the neutral gas structure is probably in 
the form of many
small dense gas clouds, and a crude upper limit is placed on the size 
of these clouds. In summary, the argument is this: any 
given $\sim 10$Mpc region of the early universe must have a baryon density 
that is close to the average for the whole universe, as discussed in 
Section~\ref{mass}. If these baryons were spread uniformly
throughout the region, they would be 
highly ionised by the UV background radiation and no absorption
would be seen. The baryons must therefore be confined into dense
clouds, occupying a small fraction of the total region.
The density must be high enough that the recombination rate balances the
photoionisation by UV background photons.
The region could contain a small number of large dense clouds, or a large
number of small dense clouds. Only in the latter case, however, would most
QSO sight-lines through the region intercept one of these clouds (smaller
clouds having a greater ratio of surface area to volume).
We now consider this argument in detail.

As discussed in Section~\ref{mass}, the density of gas within the $\sim 10$
Mpc scale volume surrounding the galaxy cluster can at most be comparable to 
$\Omega_{\rm baryon}$. If these baryons 
were spread uniformly throughout this volume, and exposed to the average
UV background at this redshift, they would be highly ionised. Why then do
the QSO sight-lines show such large neutral column? The mean free path of
 UV photons at this
redshift is $\sim 500$ Mpc (eg. Haardt \& Madau 1996): this is far greater than
the average separation of UV sources, implying that the UV background
intensity is spatially very uniform (Zuo 1992a,b, Fardal \& Shull 1993). 
In this case, we know that at least
five QSOs lie within 500 Mpc of the cluster, so if their emission is 
isotropic, lack of a UV background 
cannot explain the neutrality of the gas.

Let us
therefore consider a model of the neutral gas structure which contains at most
this average density of baryons. This mass, instead of being distributed
uniformly, is confined into clouds of scale-length $r$ and density 
$\rho$. If the clouds are sufficiently large and dense, the gas within 
them will become neutral.

For any given density, a gas cloud must have a certain minimum size for
the hydrogen within it to be neutral. This size was estimated
analytically, using standard equilibrium photoionisation, and by using 
the MAPPINGS II photoionisation code (Sutherland \& Dopita 1993). If
a UV background with a plausible spectrum and intensity is assumed
(the details make little difference to the final result), then the UV
background ionises a layer of thickness $r$ on the gas surface. The
recombination rate within this surface layer, which is proportional
to its density $\rho$ squared, must balance the photoionisation rate:
inserting numbers, we find that 
\begin{equation}
r \sim k_0 \rho^{-2},
\end{equation}
where $k_0 \sim 3 {\rm pc\ cm}^{-6}$. Unless a cloud is thicker
than this, its neutral column density will be low
(eg. Lanzetta 1991).

This gives one constraint on the size of the absorbing clouds. A second
constraint comes from our assumption that the three QSO sight-lines are
representative of the region around the galaxy cluster. This assumption 
requires that a fraction $f$ (of order unity) of all
lines of sight passing close to the cluster intersect a cloud. To keep the
discussion more general, let us allow for the possibility of sheet-like 
clouds of 
thickness $r$ and face-on cross-sectional area $A$, where $A > r^2$, so
that most of the incident UV flux enters through their face. If there are
$N$ clouds per unit volume throughout the region surrounding the galaxy 
cluster, and the region 
has a typical thickness $T$, then this condition implies that
\begin{equation}
NAT \sim 1,
\end{equation}

We can also constrain $N$, $r$ and $\rho$ by requiring that the total 
density $N r A \rho$ be at most comparable to $\rho_{\rm \max}$, the average 
baryon density of the universe at this redshift. These constraints can
only be met if $r$ is small: solving, we find that $A$
cancels, leaving us with the limits
\begin{equation}
r < \frac{\rho^2_{\rm max} T^2}{k_0} \sim 1 {\rm kpc}
\end{equation}
and
\begin{equation}
\rho > \frac{k_0}{\rho_{\rm max}T} \sim 0.03 {\rm cm}^{-3}
\end{equation}

If the absorption is caused by approximately 
spherical clouds (ie. $A \sim r^2$) these clouds could have
masses of $\sim 10^6 M_{\sun}$ or less, and a space density of 
$N > 10^5 {\rm Mpc}^{-3}$. If they are flattened, they could be
more massive and rarer.

\subsection{Argument from Metal-line Ratios\label{met}}

Most high column density QSO absorption-line systems, when observed with 
sufficiently high spectral resolution, break down into multiple components 
with different ionisation states (eg. Prochaska \& Wolfe 1997). The 
wavelength shifts between different metal-line 
species seen in our data suggest that the same is 
true for our Lyman-limit systems.

Our spectra are not of sufficient resolution to resolve this substructure, so
we are forced to use single-cloud modelling. Nonetheless, useful constraints
can be put on the cloud size. We used the MAPPINGS II photoionisation code,
as before, to estimate absorption column densities as a function of
cloud density and metallicity.

No useful constraints could be placed on the metallicity of the
gas, other than noting that it contains metals. The presence of
strong high ionisation lines, particularly C IV and Si III,
however, implies that a large part of the cloud mass cannot be at 
densities significantly higher than  $\sim 0.1 {\rm cm}^{-3}$. The 
Si III measurement could be contaminated by Ly$\alpha$ forest lines, 
but C IV should  be reliable.

Given this maximum density, the UV background ionises a layer $\sim 1$ 
kpc deep into the cloud surface. This places a lower limit on the 
scale-length 
$r$ of the absorbing clouds, if the high- and low-ionisation lines come
from the same clouds. This limit is only marginally consistent with the
upper limit on cloud size placed in Section~\ref{baryon}.

Alternatively, the absorption-line systems could be a mixture of
these low density clouds, and much smaller, higher density neutral clouds
that contribute most of the low ionisation and Ly$\alpha$ absorption.

\subsection{Argument from the Velocity Dispersions}

\begin{figure}
\psfig{file=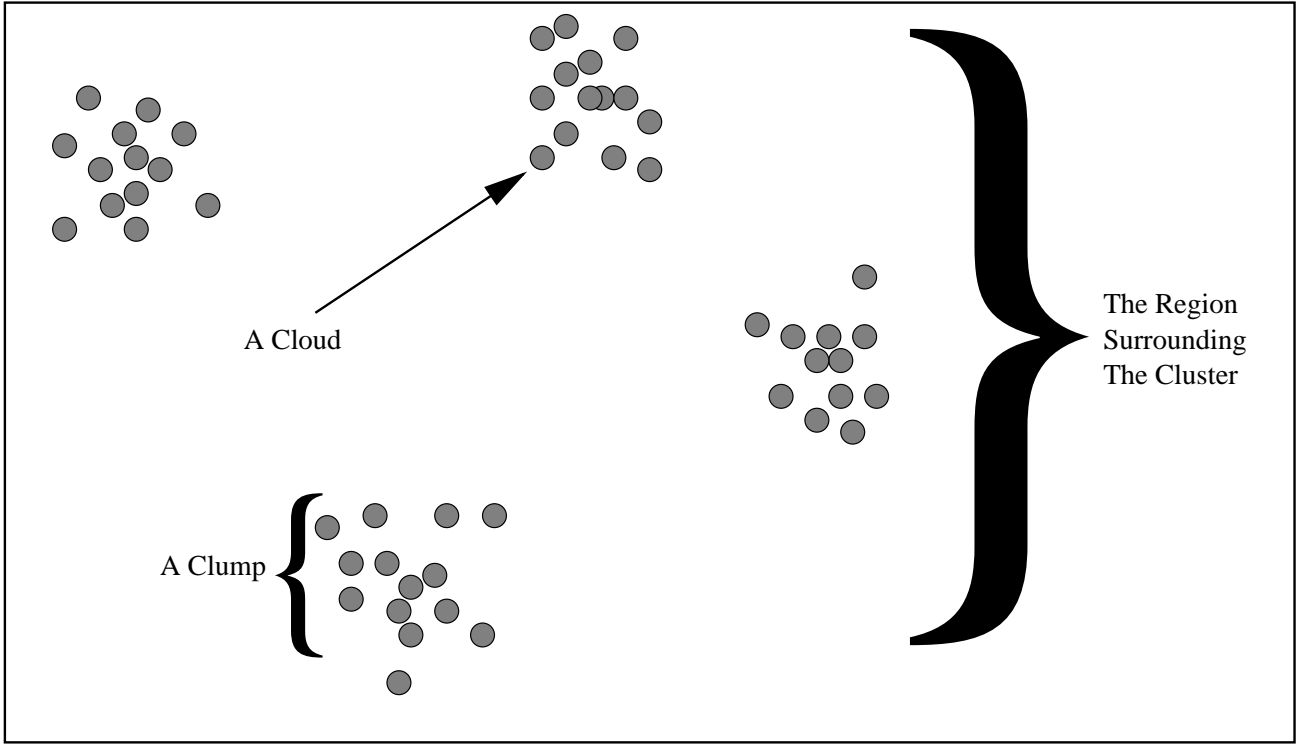}
\caption{Diagram illustrating our terminology\label{clumps}}
\end{figure}

Are the hypothesised absorbing clouds spread uniformly throughout the 
region around the galaxy cluster,
or are they gathered into clumps (Fig~\ref{clumps})? The velocity structure 
of the 
absorption-lines suggests that the latter is true. The absorption 
(Figure~\ref{threeabs}) consists of a number of discrete components: the 
velocity
width of each component ($b \sim 100 {\rm km\ s}^{-1}$) is much smaller than
the velocity dispersion between the different components
($\sim 6000 {\rm km\ s}^{-1}$). 

Each of the individual Ly$\alpha$ absorption components can only be
well fit if the absorbing gas has a velocity dispersion of $b \sim 100 
{\rm km\ s}^{-1}$ (if a velocity dispersion of $b < 50 {\rm km\ s}^{-1}$
is used in the fits, the column density has to be very large to match
the width of the base of the absorption component, in which case the
predicted damping wings are much larger than observed).
The velocity dispersions of each component, while smaller than those of
the cluster as a whole, are thus 
greater than the probable thermal velocity widths of the clouds themselves
($b < 30 {\rm km\ s}^{-1}$ for a hot photoionised phase). 

Why are the individual absorption components so broad? As discussed
in Section~\ref{met}, each component is probably made up of many
small absorbing clouds. If these small clouds are 
not gravitationally bound, the velocity dispersions 
could be Hubble-flow redshift differences between the clouds at the front 
and back of the clump: that would imply clump sizes of $\sim 100 {\rm kpc}$.
If, however, they are gravitationally bound, we can use the virial 
theorem to constrain the mass of each clump, and hence put a lower
limit on the size of each clump by requiring that the total mass of all
the clumps in the cluster not exceed a cosmologically plausible amount.

Assume that each clump has size $l$ and mass $M_c$.
Given the observed velocity dispersions $\sigma_v$, the total mass of each 
clump can be estimated from the virial theorem, and is proportional to the 
clump size:
\begin{equation}
M_c \sim \frac{l \sigma_v^2}{G}.
\end{equation}
As in Section~\ref{baryon}, we can constrain the size $l$ and space density
$N_c$ of these clumps by requiring that most sight-lines through the cluster
must intersect a clump: ie. that $f \sim 1$. Thus
\begin{equation}
N_cl^2T \sim 1.
\end{equation}
In addition, the total mass of all the clumps within a given volume, 
$N_c M_c$, can at most be comparable to the critical density of the 
universe $\rho_{\rm crit}$, as discussed in Section~\ref{mass}. Thus the 
total mass in all the clouds must be inversely proportional to the
typical cloud size $l$. Solving, we find that
\begin{equation}
l>\frac{\sigma_v^2}{GT\rho_{\rm crit}} \sim 20 {\rm kpc}.
\end{equation}
If the clumps were smaller, each would be less massive,
but so many would be needed to intersect most QSO sight-lines that the
total mass of all the clumps would be physically implausible. 

Thus each absorption component probably consists of many small
absorbing clouds gathered into clumps of size $> 20$ kpc.

\subsection{Discussion}

So, if we assume that most sight-lines passing close to the 2142$-$4420 cluster
intersect a cloud with $N_H \sim 10^{19}{\rm cm}^{-2}$, and that this
cluster is typical of high redshift clusters, these clusters must
be embedded in $\sim 10$ co-moving Mpc structures of neutral gas, made
up of large numbers of small dense gas clouds, gathered into clumps.

If these clumps exist, what could they be?
Each clump could be the halo or extended disk 
of a galaxy (eg. Prochaska \& Wolfe 1997): if they are $\sim 100$ kpc in 
size, the 
virial mass of each halo would be $\sim 10^{11} M_{\sun}$, and the whole 
cluster halo would have to contain a few hundred of these galaxy halos.
Alternatively, the velocity dispersions could represent infall into the
potential wells of protogalaxies (eg. Rauch, Haehnelt \& Steinmetz 1997). 
The properties
we infer for the absorbing gas within the cluster are strikingly
similar to the theoretical predictions of lumpy gas infalling along
filaments into protogalaxies (eg. Mo \& Miralda-Escud\'e 1996).

If the halo model is correct, we should see galaxies associated with the
absorption systems. For the Lyman-limit absorber in the spectrum of QSO 
2139$-$4434, we do see a galaxy that could plausibly be surrounded by a 
halo that is causing the absorption: one of the strong Ly$\alpha$ sources 
lies at the same redshift as the
absorber, 20$^{\prime \prime}$ ($\sim 100$ projected kpc) from the 
QSO sight-line. For the damped systems in
QSO 2139$-$4433 and QSO 2138$-$4427, however, no galaxies are seen within 
500 projected kpc of
the sight-lines. This is not however conclusive, as any galaxies could
easily lack Ly$\alpha$ emission signatures, or lie below our flux limits.

When observed at high angular resolution, many radio galaxies are seen
to be surrounded by numerous kpc-sized knots of emission, sometimes made
visible by reprocessing nuclear emission (eg. 
Stockton, Canalizo \& Ridgway 1999, Pascarelle et al. 1996). 
These knots may be the same compact gas clouds we are hypothesising, based on
the absorption. 

Could all damped Ly$\alpha$ systems be caused by absorption in galaxy
protoclusters? The space density of Lyman-break galaxy concentrations
(Steidel et al. 1998) is comparable to that of damped Ly$\alpha$ systems, 
so if all
these concentrations are optically thick throughout in Ly$\alpha$, then
most damped Ly$\alpha$ systems could indeed arise in proto-clusters.
Note, however, that our lower limit on the size of these clumps of small
absorbing clouds is greater than the tentative upper limit placed on the
size of two  damped Ly$\alpha$ systems by M\o ller \& Warren (1998) and Fynbo, 
M\o ller \& Warren (1999).

What could be the physical origin of the absorbing clouds? Thermal
instabilities can produce small cold dense clouds within the halos
of high redshift galaxies (eg. Viegas, Fria\c{c}a \& Gruenwald 1999) or 
in collapsing
proto-galaxies (Fall \& Rees 1985): it is intriguing that our inferred cloud
masses are comparable to those of globular clusters. The sound crossing
times of these small clouds would be small, so if they are long-lived
objects, they would need to be confined in some way. 

Given that all sight-lines though a clump intersect multiple absorbing 
clouds, then mean time between cloud-cloud collisions would be less than the 
time it takes for clouds to cross the clump: ie. 
100kpc/$100 {\rm km\ s}^{-1} \sim 10^9$ years (cf. McDonald \& 
Miralda-Escud\'e 1999);
further evidence for the transient nature of these clouds. What happens
to the gas in the clouds during a collision? The collision time-scale
is $\sim 1 {\rm kpc}/100 {\rm km\ s}^{-1} \sim 10^7$ years: if all the gas
were converted into stars, this would imply a star formation rate of $\sim
0.1 M_{\sun}{\rm yr}^{-1}$, which is slightly below the detection threshold
of most current surveys for high redshift star forming galaxies. Over the
neutral gas structure as a whole, however, the integrated effect of all the 
cloud-cloud collisions, if they all form stars, is a star-formation rate of
$\sim 10^3 M_{\sun}{\rm yr}^{-1}$, which should produce a diffuse
H$\alpha$ flux of $\sim 3 \times 10^{-19}{\rm erg\ cm}^{-2}{\rm s}^{-1}
{\rm arcsec}^{-2}$ (Kennicutt 1983) over the entirety of the cluster, and in 
the absence of self-absorption, a comparable flux of diffuse Ly$\alpha$
emission. We address the detection of such faint Ly$\alpha$ fluxes in
Section~\ref{diffuse}.

\section{Diffuse Ly$\alpha$ Emission\label{diffuse}}

How can we observationally determine whether neutral gas structures like the 
ones hypothesised in Section~\ref{dist} exist? One approach would be to find
larger samples of high redshift galaxy clusters with background QSOs, but
this would be enormously expensive in telescope time. Another approach is to
search for diffuse Ly$\alpha$ emission from the hypothesised neutral gas.

Even in the absence of photoionisation from young stars and/or AGN, 
the strong UV background in the high redshift universe should photoionise the
outer layers of any neutral gas clouds in the early universe, and hence induce
Ly$\alpha$ emission. This emission has been modelled by Binette et al. 
(1993) and by Gould \& Weinberg (1996) (see also Hogan \& Weymann 1987).
Predicted Ly$\alpha$ surface brightnesses are very low: typically
$10^{-19}{\rm erg\ cm}^{-2}{\rm s}^{-1}{\rm arcsec}^{-2}$ or less.
Nonetheless, if the emission covers a large area on the sky, the sensitivity 
can be increased by $\sqrt{a}$, where $a$ is the area integrated over. This 
approach has been applied to emission from individual Lyman limit systems,
and stringent upper limits placed on the surface brightness (Bunker,
Marleau \& Graham 1998).

We used results from these papers, together with our own modelling using
MAPPINGS II, to predict the diffuse Ly$\alpha$ flux from the hypothesised
neutral gas structure around the 2142$-$4420 cluster. The biggest sources of 
uncertainty are the strength of the UV ionising background at this
redshift (literature values disagree by a factor of $\sim 3$), the 
spectrum of the UV background, and the geometry of the structure (which
can also introduce a factor of $\sim 3$ uncertainty into the predicted
diffuse flux). All predictions ignore dust obscuration: if any dust is
present, then the Ly$\alpha$ fluxes will probably be reduced by a
large factor.
We predicted that the diffuse flux per unit area $F$ should lie in the range
$10^{-17} > F > 5\times 10^{-20}
{\rm erg\ cm}^{-2}{\rm s}^{-1}{\rm arcsec}^{-2}$.

\subsection{Observations}

We have made two attempts to detect diffuse Ly$\alpha$ emission from the
hypothesised neutral gas structure surrounding the 2142$-$4420 cluster.

The first attempt used the Double Beam Spectrograph on the Siding Spring 
2.3m telescope. A long-slit spectrum was taken, with the slit centred
on QSO 2139$-$4434, and the slit angle positioned so that it included
QSO 2139$-$4433. The slit was 2$^{\prime \prime}$ wide and 3$^{\prime}$
long, giving a total sky coverage of 360 square arcsec. Spectral
resolution (2 pixels) was 1.1\AA . 1800 second exposures of the cluster
field were alternated with equal length exposures on 
randomly chosen control fields one degree away (in random directions).

A total on-field exposure time of 23,400 seconds was obtained, in 
dark photometric conditions, on the nights of 9th and 10th July 1999. The
total control field exposure was identical. The night sky spectrum
was found to be very stable throughout both nights, so on- and off-field
exposures were combined, using inverse variance weighting to minimise
sky noise. The
night sky brightness at around 4110\AA\ (the expected wavelength of Ly$\alpha$
at the cluster redshift) 
 was $\sim 7 \times 10^{-18} {\rm erg\ cm}^{-2}
{\rm \AA }^{-1}{\rm arcsec}^{-2}$.

No significant excess flux was seen at wavelengths of around 411 nm in
the on-field image as compared to the off-field image. This enables us to
place a $3\sigma$ upper limit on the diffuse emission from the cluster halo, 
if it fills the slit, of $1.8 \times 10^{-18} {\rm erg\ cm}^{-2}
{\rm s}^{-1}{\rm arcsec}^{-2}$.

The second attempted measurement was made with the Taurus Tunable Filter (TTF),
on the Anglo-Australian Telescope. TTF is an imaging Fabry-Perot etalon 
system, with  a resolution at 4100\AA\ of down to 4\AA , and a monochromatic 
field of view 
approximately 5$^{\prime}$ in diameter. All observations were taken in
photometric conditions on the nights of 11th and 12th August 1999.

Charge shuffling was used to
alternate, every 60 seconds, between images of the cluster field at the
expected wavelength of Ly$\alpha$, and at a control wavelength 50 \AA\  to the
red. Observations were made at 4\AA\ resolution, with central wavelengths
of 4110 \AA\ and 4114 \AA . On-wavelength exposure times were 3000 and
3600 seconds respectively. The field of view was centred on the
brightest of the Ly$\alpha$ emitting galaxies: at  
21:42:27.48$-$44:20:28.4 (J2000).

The difference images (on-wavelength minus off-wavelength) were very
clean, with all continuum sources removed to high precision.  These images
were aligned and co-added, using inverse variance weighting. As no control 
fields were observed, we have no sensitivity to diffuse flux filling the 
field of view. We are sensitive to variations in the diffuse flux on smaller
angular scales. On scales of $20^{\prime \prime} \times 20^{\prime \prime}$,
we see no regions of excess flux, to a 
$3 \sigma$ upper limit of $5 \times 10^{-19} {\rm erg\ cm}^{-2}
{\rm s}^{-1}{\rm arcsec}^{-2}$, at both wavelengths.

\subsection{Summary}

We failed to detect diffuse emission from any neutral gas around the 
2142$-$4420 cluster. Our observations were not, however, deep enough
to rule out the presence of even dust-free neutral gas. They do, however,
demonstrate the darkness and stability of the sky background at these
blue wavelengths.

\section{Conclusions\label{conclusions}}

The QSO absorption-line data on the 2142$-$4420 cluster are provocative.
If taken at face value, they suggest that this high redshift cluster,
and perhaps all of them, are surrounded by  $\sim 5$ Mpc structures 
of neutral hydrogen, of total mass $\sim 10^{12} M_{\sun}$. This gas
would be gathered into a series of $> 20 $kpc clumps, each of which is made
up of hundreds of small ($< 500$ pc), dense ($> 0.1 {\rm cm}^{-3}$)
clouds. These clouds may merge to form cluster galaxies, collapse to form
globular clusters, or perhaps dissolve into X-ray intra-cluster medium
by redshift zero.

Many more observations of neutral gas around high redshift
clusters will be needed to verify this picture.

\section*{Acknowledgements}

We wish to thank Phillippe V\'eron and Mike Hawkins for making their
spectrum of QSO 2139$-$4433 available to us, Ralph Sutherland for
his help and advice with the photoionisation modelling, Mark Phillips
for taking the CTIO spectrum for us, Catherine Drake and Joss Bland-Hawthorn
for their help with the TTF observations and Luc Binette, Gerry Williger and 
Povilas Palunas for helpful discussions.

\section*{References}

\reference  Adelberger, K.L., Steidel, C.C.,
Giavalisco, M., Dickinson, M., Pettini, M. \& Kellogg, M. 1998, ApJ, 505, 18

\reference  Bagla, J.S. 1998, MNRAS, 299, 417

\reference  Baugh, C.M., Benson, A.J., Cole, S.,
Frenk, C.S. \& Lacey, C.G. 1999, MNRAS, 305, L21

\reference Bicknell, G.V., Sutherland, R.S., van Breughel, W.J.M.,
Dopita, M.A., Dey, A. \& Miley, G.K. 2000, ApJ, submitted.

\reference Binette, L., Wang, J.C.L., Zuo, L., \& Magris, C.G. 1993, AJ, 105,
797

\reference Binette, L., Kurk, J.D., Villar-Martin, M. \&  
R\"ottgering, H.J.A. 2000, A \& A, submitted.

\reference  Brainerd, T. G., \& 
Villumsen, J. S. 1994, ApJ, 431, 477

\reference Bunker, A.J., Marleau, F.R., \& Graham, J.R. 1998, ApJ, 116,
2086

\reference  Campos, A., Yahil, A., Windhorst, 
R. A., Richards, E. A.,
Pascarelle, S., Impey, C. \& Petry, C., 1999, ApJL , 511, L1

\reference Carilli, C.L., Harris, D.E., Pentericci, L., R\"ottgering, H.J.A.,
Miley, G.K. \& Bremer, M.N. 1998, ApJL, 494, 143

\reference Carilli, C.L.,  R\"ottgering, H.J.A., van Ojik, R.,
Miley, G.K. \& van Breughel, W.J.M. 1997, ApJS, 109, 1

\reference  Cen, R. 1998, ApJ, 509, 16

\reference  Cen, R. \& Ostriker, J.P. 1999, ApJ,  514, 1

\reference  C\'olin, C. P.,
Carlberg, R. G., \& Couchman, H. M. P. 1997, ApJ, 490, 1

\reference  Colless, M., Ellis, R. S., Taylor, K.
\& Hook, R. N. 1990, MNRAS, 244, 408

\reference  Davis, M. \& Peebles, P.J.E.
1983, ApJ, 267, 465

\reference  Djorgovski, S.G., Odewahn, S.C.,
Gal, R.R., Brunner, R. \& Carvalho, R.R. 1999, to appear in ``Photometric
Redshifts and the Detection of High Redshift Galaxies'', ed. Weymann, R.J.,
Storrie-Lombardi, L., Sawicki, M. \& Brunner, R. (San Francisco: ASP
conference Series) (astro-ph/9908142)

\reference  Donahue, M. \& Voit, G. M. 1999,
ApJL, in press (astro-ph/9907333)

\reference  Drinkwater, M. J., 
Barnes, D. G. \& Ellison, S. L.
1995, PASA, 12, 248

\reference  Deltorn, J.-M, Le Fevre, O.,
Crampton, D. \& Dickinson, M. 1997, ApJ, 483, 21

\reference  Fall, S. M. \& Rees, M. J. 1985, ApJ, 
298, 18

\reference  Fardal, M.A. \& Shull, J.M. 1993,
ApJ, 415, 524

\reference  Francis, P. J., \& Hewett, P. C. 
1993, AJ , 105, 1633

\reference  Francis, P. J., et al. 1996, ApJ, 457, 
490 

\reference  Francis, P. J., Woodgate, 
B. E. \& Danks, A. C., 1997,
ApJL, 482, 25

\reference  Fry, J. N. 1996, ApJ, 461, 65

\reference  Fynbo, J. U.,
M\o ller, P. \& Warren, S. J. 1999, MNRAS, 305, 849

\reference  Giavalisco, M., Steidel, C.C.,
Adelberger, K.L., Dickinson, M.E., Pettini, M. \& Kellogg, M. 1998, ApJ, 
503, 543

\reference Gould, A., \& Weinberg, D.H.1996, ApJ, 468, 462

\reference  Groth, J.E. \& Peebles, P.J.E.
1977, ApJ, 217, 385

\reference  Haardt, F. \& Madau, P. 1996,
ApJ, 461, 20

\reference  Heisler, J., Hogan, C. J., 
\& White, S. D. M. 1989, ApJ,
347, 52

\reference Hogan, C.J., \& Weymann, R.J. 1987, MNRAS, 225, 1P

\reference  Hu, E.M., Cowie, L. L., 
\& McMahon, R. G. 1998, ApJL, 502, L99

\reference Ivison, R.J., Dunlop, J.S., Smail, I., Dey, A., Liu, M.C.
\& Graham, J.R. 2000, ApJ in press (astro-ph/0005234)

\reference  Jenkins, A., Frenk, C.S., 
Pearce, F.R., Thomas, P.A., Colberg, J.M., White, S.D.M., Couchman, H.M.P.,
Peacock, J.A., Efstathiou, G. \& Nelson, A.H. 1998, ApJ, 499, 20 

\reference  Kennicutt, R.C., jr. 1983, ApJ, 272, 54

\reference  Kolb, E. M., \& Turner, M. S., 
1990, ``The Early Universe'',
(Addison-Wesley, New York), 335

\reference  Lanzetta, K.M. 1991, ApJ, 375, 1

\reference  Ma, C-P. 1999, ApJ, 510 32

\reference  Malkan, M. A.,
Teplitz, H. \& McLean, I.S. 1996, ApJL, 468, L9

\reference  Mar, D. P. \& Bailey, G. 1995, PASA, 12 
239 

\reference 
Mart\'{\i}nez-Gonz\'alez, E., Gonz\'alez-Serrano, J. I.,
Cay\'on, L., Sanz, J. L., \& Mart\'{\i}n-Mirones, J. M. 1995, A \& A, 303, 379

\reference 
Matarrese, S., Coles, P., Lucchin, F. \& Moscardini, L. 1997, MNRAS, 286, 
115

\reference  McDonald, P. \&
Miralda-Escud\'e, J. 1999, ApJ, 519, 486

\reference  Mo, H.J. \&
Miralda-Escud\'e, J. 1996, ApJ, 471, 582

\reference  M\o ller, P. \& Warren, S.J.
1998, MNRAS, 299, 661

\reference  Moscardini, L., Coles, P.,
Lucchin, F., Matarese, S. 1998, MNRAS, 299, 95

\reference  Pascarelle, S.M., Windhorst, R.A.,
Keel, W.C. \& Odewahn, S.C. 1996, Nature, 383, 45

\reference  Pascarelle, S.M.,
Windhorst, R.A. \& Keel, W.C. 1998, AJ, 116, 2659

\reference Pentericci, L., R\"ottgering, H.J.A., Miley, G.K.,
Carilli, C.L. \& McCarthy, P. 1997, A \& A, 326, 580

\reference Pentericci, Van Reeven, L.W., Carilli, C.L., R\"ottgering, H.J.A.,
\& Miley, G.K. 2000, A \& A, in press (astro-ph/0005524)

\reference  Porciani, C. 1997, MNRAS, 290, 639

\reference  Prochaska, J. X. \& Wolfe, 
A. M. 1997, ApJ, 487, 73

\reference  Quashnock, J. M., 
Vanden Berk, D. E., \& York, D. G. 1996, ApJL, 472, L69

\reference  Rauch, M., Haehnelt, 
M. G. \& Steinmetz, M. 1997, ApJ,
481, 601

\reference  Renzini, A. 1997, ApJ, 488, 35

\reference  Rosati, P., Della Ceca, R.,
Norman, C. \& Giacconi, R. 1998, ApJL, 492, L21

\reference  Steidel, C. C. 1990, ApJS, 74, 37

\reference Steidel, C.C., Adelberger, K.L., Shapley, A.E., Pettini, M.,
Dickinson, M. \& Giavalisco, M. 2000, ApJ 532, 170

\reference  Steidel, C. C., Giavalisco, M., 
Pettini, M., Dickinson, M. \&
Adelberger, K. L. 1996, ApJL, 462, L17

\reference  Steidel, C. C., Adelberger, K. L., 
Dickinson, M., 
Giavalisco, M., Pettini, M., \& Kellogg, M. 1998, ApJ, 492, 428

\reference  Stockton, A.,
Canalizo, G. \& Ridgway, S.E. 1999, ApJL, 519, 131

\reference  Sutherland, R. S. \& Dopita, 
M. A. 1993, ApJS, 88, 253

\reference  Tegmark, M. \& Peebles, P.J.E.
1998, ApJL, 500, L79

\reference  V\'eron, P. \& Hawkins, M. R. S. 
1995, A \& A, 296, 665

\reference  Viegas, S. M.,
Fria\c{c}a, A.C.S. \& Gruenwald, R. 1999, MNRAS, 309, 355

\reference  Walker, T. P., Steigman, G., 
Kang, H-S., Schramm, D. M.
\& Olive, K. A. 1991, ApJ, 376, 51

\reference  Zuo, L. 1992a, MNRAS, 258, 36

\reference  Zuo, L. 1992b, MNRAS, 258, 45

\end{document}